\documentstyle[12pt]{article}
\date{}

\begin{document}
\title{Collective excitation frequencies of a Bose -- Einstein 
condensate with electromagnetically induced $1/r$ attraction}
\author{I.E.Mazets,\\Ioffe Physico-Tecnical Institute, 194021 
St.Petersburg, Russia, \\e-mail: mazets@astro.ioffe.rssi.ru}
\maketitle 

\begin{abstract}
We investigate collective excitations of an  atomic 
Bose-Einstein condensate in a recently discovered 
regime [D.~O'Dell et al., Phys. Rev. Lett. {\bf 84}, 5687 (2000)] 
of a balance between electromagnetically induced $1/r$ attraction 
and a short-range interatomic repulsion. The corresponding 
frequencies for monopole, dipole, and quadrupole modes are 
calculated numerically in the  zero 
temperature case. 
\end{abstract}

\vskip 18pt

Theoretical as well as experimental studies of dilute atomic 
Bose -- Einstein condensates (BECs) have made an impressive progress 
since 1995, when quantum degeneration in alkali metal 
wapours was reached at the first time \cite{s1995}. These systems 
are interesting especially due to a presence of an interaction 
between atoms weak enough to provide a development of quite 
precise theories describing the properties of these non-ideal 
degenerate Bose-gases. The interatomic potentials act on a 
short range, moreover, at ultralow temperatures they are effectively 
described by contact pseudo-potentials containing a single 
parameter, the $s$-wave  scattering length $a$ \cite{s1999}. 
Here we consider $a>0$, i.e. the short-range potential 
provides mutual repulsion of atoms.

Recently, a new kind of atomic BEC has been proposed in 
Ref.\cite{r1}. Namely, it is shown that the particular 
configurqation of  intense off-resonant laser beams gives rise to 
an effective $1/r$ interatomic attraction between atoms located 
well within the laser wavelength (since inelastic collisions 
destroy very rapidly BECs of number densities more than 
$10^{14}$ cm$^{-3}$, a long-wavelength 
infrared laser has to be considered). 
The directions and polarizations of different incident laser 
beams can be chosen in such a way that the static (near-zone) 
$1/r^3$ component of dipole-dipole interaction energy is 
effectively ''averaged out'', and only  $1/r$ attraction remains. 
Finally, the two-particle interaction energy including both usual 
short-range interactions and such an ''electromagnetically induced 
gravitation'' reads as 
\begin{equation}
U_{12}({\bf r})=\frac {4\pi \hbar ^2 a}M \delta ({\bf r})-\frac ur   ,
\label{au}\end{equation}
where $M$ is the atomic mass and the long-range interaction 
constant $u$ is expressed in SI units as \cite{r1}
$$
u=\frac {11 Iq^2\alpha ^2}{4\pi \epsilon _0^2c}.
$$
The laser beam intensity and wavenumber are denoted by $I$ and $q$, 
respectively; $\alpha $ is the atomic polarizability at the 
frequency $q/c$ (practically, because the external radiation 
belongs to the IR range, $\alpha $ is equal to the static 
polaqrizability). A useful parameter $a_*=\hbar ^2/(Mu)$ can be 
called the Bohr radius for the gravitation-like potential. 
Practically, the ''gravitation'' is weak enough, so always $a_*\gg a$.  

In Ref.\cite{r1} it is shown that a self-bound solution of the 
modified Gross -- Pitaevskii equation taking into account the 
gravitation-like interatomic potential exists. A cloud of atoms 
located in a radiation field of necessary geometry giving rise to 
$1/r$ attraction is stable without 
any confining external potential of magnetic or optical trap. 
The situation is similar to that of astrophysics where we have 
stars --- objects those are stable due to 
a balance between gravitation and a Boltzmannian gas 
pressure. In our case the pressure arises from the two sources: 
interatomic repulsion due to the short-range potential and the 
''quantum pressure'' corresponding to the kinetic energy term in the 
Schr\"odinger equation. 

Let us consider a zero-temperature case. 
Expressing the BEC order parameter through condensate density 
$n$ and phase $\phi $ as usual \cite{s1999}, 
$\Psi =n\exp(i\phi )$, we rewrite the complex 
Gross -- Pitaevskii equation as a set of two real equations: 
\begin{equation}
\frac {\partial n}{\partial t}+\nabla (n{\bf v})=0 ,  \label{twoeq}
\end{equation}
$$
M\frac {\partial {\bf v}}{\partial t}+\nabla \left( \frac 12 Mv^2
-\frac {\hbar ^2}{2m\sqrt{n}}\nabla ^2\sqrt{n}+
\frac {4\pi \hbar ^2 a}m n +U_g\right)   =0, 
$$
where the local ''gravitation potential'' is 
$$
U_g({\bf r})=-u \int d^3{\bf r}^\prime \, 
\frac {n({\bf r}^\prime )}{\left| {\bf r}-{\bf r}^\prime \right| }. 
$$
The velocity field is introduced by the relation 
${\bf v}=\hbar  \nabla \pi /M$. 

If $N\gg \sqrt{a_*/a}$ ($N$ is the total number of atoms in the BEC) 
then the set of Eqs.(\ref{twoeq}) admits an analytical 
steady state solution \cite{r1}. 
Indeed, in such a limiting case the kinetic 
energy term is small and can be omitted; the ''gravitation'' is 
balanced by the $s$-wave scattering. The density of such a self-bound 
degenerate bosonic cloud is equal to 
\begin{equation}
n_0({\bf r})=\frac N{4R_0^2}\frac {\sin (\pi r/R_0)}r \theta (R_0-r) 
=0,  \label{n0}  
\end{equation}
where $\theta $ is the Heavyside function, $R_0=\pi \sqrt{a_*a}$ is 
the radius of the cloud boundary (in Ref.\cite{r1} there is an  
arythmetic 
error in the numerical coefficient in the expression for $R_0$). 

In the present paper, we calculate the eigenfrequencies of 
oscillations of such a self-bound cloud. We introduce small 
perturbations $n_1$ and ${\bf v_1}$ of the density and velocity, 
respectively: $n=n_0+n_1$, ${\bf v}={\bf v}_1$. 
procedure of linearizing of Eqs.(\ref{twoeq}) is standard 
\cite{str}. We choose the time dependence of perturbation in the 
form of $\exp (-i\omega t)$. 
We calculate the eigenfrequenciues $\omega $ within the Thomas -- Fermi 
approximation, i.e., we  neglect  
the kinetic energy term in the equation for the velocity, 
After elimination of the variable ${\bf v}_1$, we get finally 
\begin{equation}
\tilde \omega ^2 n_1(\tilde{\bf r})+
\tilde \nabla \left[ \frac {\sin (\tilde{r})}{\tilde r}\tilde \nabla 
\left( n_1(\tilde {\bf r})-
\frac 1{4\pi } \int d^3 \tilde{\bf r}^\prime \,  
\frac {n_1({\bf r}^\prime )}{\left| \tilde{\bf r}-
\tilde{\bf r}^\prime \right| }
\right)  \right]  =0 .         \label{oneeq}
\end{equation}
Eq.(\ref{oneeq}) is written using dimensionless quantities 
\begin{equation} 
\tilde {\bf r}=\frac \pi {R_0}{\bf r}, \quad 
\tilde \omega =\omega \sqrt{\frac {R_0^5M^2}{\pi ^4N\hbar ^2a}} .
\label{dimless}
\end{equation}
It is convenient to introduce the ''gravitation potential 
perturbation'' variable   
\begin{equation}
\xi (\tilde {\bf r}) =
\frac 1{4\pi } \int d^3 \tilde{\bf r}^\prime \,  
\frac {n_1({\bf r}^\prime )}{\left| \tilde{\bf r}-
\tilde{\bf r}^\prime \right| }     , \label{xxx}
\end{equation}
so that $n_1=-\tilde \nabla ^2\xi $. Then Eq.(\ref{oneeq}) can be 
identically transformed to 
\begin{equation}
\tilde \omega ^2\tilde \nabla ^2 \xi +\tilde \nabla \left[ 
\frac {\sin (\tilde{r})}r\left( \tilde \nabla ^2\xi +\xi \right) 
\right]  =0.
\label{eqxi} 
\end{equation}

To solve this equation, firstly, we  note that the orbital momentum $l$ 
and its projection $m$ to an arbitrary axis are conserved. So we can set 
$$
\xi =\xi _l (\tilde{r})Y_{lm}(\vartheta, \varphi ) ,
$$
where $Y_{lm}$ is the spherical harmonic. The only remaining thing now 
is to set the boundary condition for Eq.(\ref{eqxi}). They can be 
derived from the tailoring of the inner and outer solutions at 
$\tilde{r}=\pi $ (i.e., $r=R_0$). First of all, a perturbation does not 
change the total number $N$ of atoms. So, for monopole modes,  
the radial derivative of 
the potential perturbation vanishes at  
$\tilde r>\pi $. For modes 
with $l=1,2,3,\, ...$ at the same distances $\xi _l$ decreases as 
$r^{-(l+1)}$. Hence, the boundary conditions are 
\begin{equation}
\frac {d\xi _0}{d\tilde r}\vert _{\tilde r=\pi }=0              
\label{bc0}    
\end{equation}
for the monopole mode and 
\begin{equation}
\frac 1{\xi _l}\frac {d\xi_l}{d\tilde r} 
\vert _{\tilde r=\pi }=-\frac {l+1}\pi   
\label{bcothers}
\end{equation}
for the other modes ($l=1,2,3, \, ...$). 

We solved Eq.(\ref{eqxi}) numerically by expansion in series:  
$$
\xi _l(\tilde r)=\sum _k C_{lk}j_l(\gamma _{lk}\tilde r/\pi ). 
$$
Here $j_l$ is the spherical Bessel function \cite{math}. To satisfy 
the boundary conditions, one must choose the coefficient 
$\gamma _{lk}$ equal to the $k$-th root of the equation 
\begin{equation}
j_{l^\prime }(\gamma )=0,           \label{root}
\end{equation}
where 
$$
l^\prime =1,\qquad l=0, 
$$
$$
l^\prime =l-1,\qquad l=1,2,3,\, ...\, .
$$
For $l=1$, the lowest frequency is exactly zero. 
Indeed, the lowest dipole mode corresponds to 
a translation of a self-bound BEC as a whole in a space free of any 
external trap potential. 

We calculated numerically the four lowest frequencies for 
$l=0,1,$ and 2. 
The estimated numerical uncertainty is of order of one per cent. Thus, 
$\tilde \omega $ is equal to 

$0.617,\, 1.27,\, 1.85,\, 2.39,\, ... $ for $l=0$, 

$0,\, 0.898,\, 1.51,\, 2.06,\, ... $ for $l=1$, 

$0.391,\, 1.10,\, 1.71,\, 2.28,\, ...$ for $l=2$. 

Finally, we note that calculation of quantum depletion in a BEC 
self-bound by ''electromagnetically induced gravitation'' could be 
interesting. 

This work is supported by the Russian Foundation for Basic Research 
(project 99--02--17076) and the State Program ''Universities of 
Russia'' (project 015.01.01.04).


\begin{thebibliography}{9}
\bibitem{s1995} M.H. Anderson et al., Science {\bf 269}, 198 (1995); 
C.C. Bradley et al., Phys. Rev. Lett. {\bf 75}, 1687 (1995); 
K.B. Davis et al., Phys. Rev. Lett., {\bf 75}, 3969 (1995). 

\bibitem{s1999} For a recent review of the atomic BEC theory 
see, e.g., F. Dalfovo, S. Giorgini, L.P. Pitaevskii, and 
S. Stringari, Rev. Mod. Phys. {\bf 71}, 463 (1999). 

\bibitem{r1} D. O'Dell, S. Giovanazzi, G. Kurizki, and V.M. Akulin, 
Phys. Rev. Lett. {\bf 84}, 5687 (2000). 

\bibitem{str} S. Stringari, Phys. Rev. Lett. {\bf 77}, 2360 (1996). 

\bibitem{math} M. Abramowitz and I.A. Stegun, eds. 
''{\it Handbook of Mathematical Functions}'', Nat. Bureau of 
Standards, 1964, Chapter 10. 
\end{thebibliography}
\end{document}